\documentclass[twocolumn,showpacs,preprintnumbers,amsmath,amssymb,floatfix]{revtex4}

\usepackage{graphicx}
\usepackage{dcolumn}
\usepackage{bm}

\newcommand{\beq}{\begin{equation}}
\newcommand{\eeq}{\end{equation}}
\newcommand{\bey}{\begin{eqnarray}}
\newcommand{\eey}{\end{eqnarray}}

\begin{document}

\title{Variable Equation of State for Generalised Dark Energy Model}

\author{Saibal Ray\footnote{saibal@iucaa.ernet.in}}
\affiliation{Department of Physics, Government College of Engineering 
\& Ceramic Technology, Kolkata 700 010, West Bengal, India}

\author{Farook Rahaman\footnote{farook\_rahaman@yahoo.com}}
\affiliation{Department of Mathematics, Jadavpur University, Kolkata 700 032, West Bengal, India} 

\author{Utpal Mukhopadhyay\footnote{utpal1739@gmail.com}}
\affiliation{Satyabharati Vidyapith, Barasat, North 24 Parganas, 
Kolkata 700 126, West Bengal, India}

\author{Ruby Sarkar\footnote{rubyfantacy@yahoo.co.in}}
\affiliation{Future Campus School, Sonarpur Station Road, Kolkata 700 150, West Bengal, India} 

\date{\today}

\begin{abstract}
We present a model for the present accelerating Universe and focus on the 
different important physical variables involved in the model under 
the phenomenological assumption $\Lambda \propto H^2$ with a prescription 
for equation of state parameter in the form $\omega(t)=\omega_0+\frac{\omega_1\tau}{t^2}$, 
where $\omega_0$ and $\omega_1$ are two constants and $\tau$ is a parameter having dimension 
of time $t^2$. General expressions for the density parameter $\Omega$ and deceleration parameter 
$q$ are obtained which under specific bound reproduce some of the previous results. 
We explore physical features of these parameters which (i) provide the scenario of 
complete evolution of the cosmos with $\omega(t)$ and (ii) agree mostly with the 
observational status of the present phase of the accelerating Universe.
\end{abstract}



\maketitle

\section{Introduction}

Ever since the discovery of an Universal acceleration \cite{Riess
et al.1998,Perlmutter et al.1998}, great efforts are going on for
finding a satisfactory explanation of this effect. Most of
the works related to this problem are based on the assumption that
an exotic type of energy, termed as {\it dark energy}, is the
causative agent of this acceleration although alternative
explanations are also available in the literature
\cite{Barrow1999,Garcia-Berrow1999,Celerier2000}. Phenomenological
models are also contenders in this race for understanding the
elusive dark energy. A number of phenomenological models involving
time-dependent cosmological term $\Lambda$ have already been
proposed and tested by various workers from different standpoints
(for an overview see Overduin and Cooperstock \cite{Overduin1998}
and Sahni and Starobinsky \cite{Sahni2000}). In favor of 
$\Lambda$-decay scenario, irrespective of whether
 they come from extended theories of gravity or phenomenological
 considerations, it is argued \cite{Overduin1998,Ray2007a} that 
(i) they have been shown to address  a number of pressing problems 
in cosmology; (ii) many are independently  motivated, e.g., 
by dimensional arguments, or as limiting cases of more
 complicated theories; (iii) most are simple enough that meaningful
 conclusions can be drawn about their viability and (iv) successful
implementation would point either towards the eventual Lagrangian
formulation or a more complete theory.

A very important parameter for the above dark energy investigation is
that of the equation of state parameter $\omega$. For pressureless dust,
radiation, stiff-fluid and vacuum-fluid dominated Universe, one
can assume the values $0$, $1/3$, $1$ and $-1$ respectively for
$\omega$. Some other limits obtained from observational results
coming from SN Ia data \cite{Knop2003} and SN Ia data corroborated
with CMBR anisotropy and galaxy clustering
statistics~\cite{Tegmark2004} are $-1.67<\omega<-0.62$ and
$-1.33<\omega<-0.79$ respectively. However, it is not at all
obligatory to use a constant value of $\omega$. In fact, it can be
a function of time, redshift or scale factor \cite{Chevron2000,Zhuravlev2001,Peebles2003}. 
But, due to lack of conclusive observational data which may enable us to
distinguish between a variable and constant equation of state,
usually a constant value of $\omega$ is used
\cite{Kujat2002,Bartelmann2005}. Coming to the case of particular
relationship of variable $\omega$ with redshift, it can be linear
like $\omega(z)=\omega_0+\omega'z$ where
$\omega'=(d\omega/dz)_{z=0}$ \cite{Huterer2001,Weller2002} or non-linear as
$\omega(z)=\omega_0+\omega_1z/(1+z)$ \cite{Polarski2001,Linder2003a}. 
So, as far as the scale factor 
dependence of $\omega$ is concerned, the parametrization
$\omega(a)=\omega_0+\omega_a(1-a)$, where $\omega_0$ is the
present ($a=1$) value and $\omega_a$ is a measure of the time
variation $\omega'$, is widely used in the literature
\cite{Linder2008}. Different forms of $\omega_a$ are also
available in various works \cite{Linder2003a,Rapatti2005}. Some
other models such as kink model \cite{Corasaniti2003}
and e-fold model \cite{Linder2005} have tried to
construct a monotonic $\omega(a)$ involving four parameters which
makes transition from some asymptotic past value $\omega_p$ to
some asymptotic future value $\omega_f$. Physical supports in
favor of $\omega_a$ parametrization have been given by Linder
\cite{Linder2003a,Linder2003b}. In quintessence models involving 
scalar fields give rise to time-dependent $\omega$
\cite{Ratra1988,Turner1997,Caldwell1998,Liddle1999,Steinhardt1999}.
There are also some cosmological models with a time dependent equation of state 
parameter available in the literature with Kaluza-Klein metric and wormholes 
\cite{Bhui2005,Rahaman2006,Rahaman2009}. 

Another aspect of the accelerating Universe is related with the signature flipping
of the deceleration parameter $q$. In fact, in the present dark
energy dominated Universe, the second largest constituent is
another dark component, viz. dark matter. About $25\%$ 
contribution from the matter content of the Universe is in the form
of dark matter. Now, in an ever accelerating Universe,
hierarchical structure formation is impossible. A primary
criteria, therefore, for the $\Lambda$-CDM Universe is that the present
acceleration is a recent phenomena and was preceded by a period of
deceleration. Hence the deceleration parameter $q$ must have
changed its sign sometime during cosmic evolution. Both
theoreticai \cite{Amendola2003,Padmanabhan2003} and observational
\cite{Riess2001} works support this physical idea. So, $q$ must be
time-dependent so that the transition of the Universe from a
decelerating to an accelerating phase can occur smoothly. This
point must be kept in mind while dealing with dark energy
investigation regarding present accelerating Universe.

In recent years various forms of time-dependent $\omega$ have been
used for variable $\Lambda$ models, viz. $\dot\Lambda \propto H^3$,
\cite{Mukhopadhyay2007,Mukhopadhyay2008a,Usmani2008},
$\Lambda \propto \dot H$ \cite{Mukhopadhyay2005} and
$\Lambda \propto H^2$ \cite{Mukhopadhyay2008b}. Recently Ray et al. 
\cite{Ray2007b} have shown the equivalence of the three $\Lambda$
models, viz. $\Lambda \propto H^2$, $\Lambda \propto \ddot a/a$ and
$\Lambda\propto \rho$ for constant $\omega$. It was mentioned in
that work of Ray et al.\cite{Ray2007b} that in a later work, a more accurate
analysis may be made by considering a time-varying equation of
state parameter whereas Mukhopadhyay et al. \cite{Mukhopadhyay2008b}
have already investigated the $\Lambda \propto H^2$ model for a
particular type of variable $\omega$ with the full physical range of 
$\alpha$. So, the purpose of the present paper is to re-examine the status 
of the same $\Lambda \propto H^2$ model for the presciption 
$\omega(t)=\omega_0+\frac{\omega_1\tau}{t^2}$, where $\omega_0$ and $\omega_1$ 
are two constants and $\tau$ is a parameter having dimension of $t^2$. 
The scheme of the investigation is as follows: In the Section 2 
the Einstein field equations and their solutions are provided 
whereas Section 3 deals with the physical features of different parameters. 
In Section 4 some comments are made on the basis of the results 
obtained in the investigations.

\section{The Einstein Field Equations and Their Solutions}

 The Einstein field equations are
 \begin{eqnarray}
  R^{ij} - \frac{1}{2}Rg^{ij} = -8\pi G\left[T^{ij} - \frac{\Lambda}{8\pi G}g^{ij}\right],\label{efe1}
\end{eqnarray}
where the erstwhile cosmological constant $\Lambda$ is assumed as a
function of time, viz. $\Lambda = \Lambda(t)$ and the velocity of
light $c$ in vacuum is unity when expressed in relativistic units.

In this connection it is to be mention that invariant property of 
$\Lambda$ under Lorentz transformation is not satisfied for arbitrary
systems (e.g., material systems and radiation). It has been argued that 
the energy density of vacuum represents a scalar function of the
four-dimensional space-time coordinates so that it satisfies the
Lorentz symmetry \cite{Gliner1965,Majernik2001}. On the other hand 
Vishwakarma \cite{Vishwakarma2001} considered a particular
Ricci-symmetry under the framework of general relativity which is
the contracted Ricci-collineation along the fluid flow vector and
argues that this symmetry does demand Lambda to be a function of
time (and space, in general). In this context we would also like 
to mention that by the application of renormalization group approach
of quantum field theory in cosmology Shapiro et al.\cite{Shapiro2005} 
have derived generalized energy conservation laws for
variable $\Lambda$ models. Such generalized laws have also been deduced 
by Vereschagin et al.\cite{Vereschagin2006} based on the work of 
Gurzadyan and Xue \cite{Gurzadyan2003}. However, comment made by 
Beesham \cite{Beesham1993} and Shapiro et al. \cite{Shapiro2005} 
that for simplicity, one may still retain the usual energy
conservation law even when $\Lambda$ is a variable. In the present investigation, 
therefore, we have used the same energy conservation law for solving the 
dynamical cosmological equation. 

For the spherically symmetric
Friedmann-Lema{\^i}tre-Robertson-Walker (FLRW) metric
\begin{eqnarray}
 ds^2 = -dt^2 + a(t)^2\left[\frac{dr^2}{1 -
kr^2} + r^2 (d\theta^2 + sin^2\theta d\phi^2)\right],\label{flrwm}
\end{eqnarray}
where $a$ is the scale factor and the curvature constant $k$ is
$-1, 0, +1 $ respectively for open, flat and close models of the
Universe, the Einstein field equations (\ref{efe1}) take the forms as
\begin{eqnarray} 
\left(\frac{\dot a}{a}\right)^2+\frac{k}{a^2} =
\frac{8\pi G}{3}\rho + \frac{\Lambda}{3}, \label{efe2}
\end{eqnarray}
\begin{eqnarray} 
\frac{\ddot a}{a} = - \frac{4\pi G}{3} (\rho +
3p)+ \frac{\Lambda}{3}. \label{efe3}
\end{eqnarray}

From equations (\ref{efe2}) and (\ref{efe3}), one can arrive at the equation
\cite{Ray2007b}
\begin{eqnarray}
\left(\frac{\dot a}{a}\right)^2 + \left[3\left(\frac{1+
    w}{1 + 3w}\right) - 1 \right]\frac{\ddot a}{a} + \frac{k}{a^2} =
        \left(\frac{1+ w}{1 + 3w}\right)\Lambda,\label{efe4}
\end{eqnarray}
where the time-dependent barotropic equation of state parameter
$\omega(t)$ is given by $\omega(t)= p(t)/\rho(t)$.

Let us use the {\it ansatz} $\Lambda=3\alpha H^2$. This is one of the 
very well known and widely used phenomenological models as proposed by 
Carvalho et al. \cite{Carvalho1992} and Waga\cite{Waga1993} 
from dimensional arguments. However, Lima and
Carvalho \cite{Lima1994} have considered the same model from different
point of view. According to Vishwakarma \cite{Vishwakarma2002},
one can obtain the relationship $\Lambda \propto H^2$ when the
cosmological term $\Lambda$ is expressed in terms of Planck energy
density. Moreover, it has been demonstrated by other authors 
\cite{Cohen1977,Hsu2004} that $\Lambda \propto H^2$ law can also be
deduced by using effective field theory and black hole
thermodynamics.

By the use of this {\it ansatz} $\Lambda=3\alpha H^2$ in the equation (\ref{efe4}), 
we arrive at the following form
\begin{eqnarray}
H^2+\frac{2}{(1+3\omega)}(\dot H+H^2)=
\frac{1+\omega}{(1+3\omega)}3\alpha H^2
\end{eqnarray}
which, on simplification becomes
\begin{eqnarray}
\frac{dH}{H^2}= -\frac{3(1-\alpha)}{2}(1+\omega)dt.\label{H2}
\end{eqnarray}

Now, it is important to consider a suitable form of the 
equation of state parameter $\omega$. As already mentioned 
in the Introduction that it can be time-dependent and 
even it may be a function of the red-shift $z$ or scale factor $a$ 
as well. For time-dependence Mukhopadhyay et al. \cite{Mukhopadhyay2010} 
suggest it in the form $\omega(t)= \omega_0+\omega_1 t^n$ 
which can be regarded as a generalization of the special form 
$\omega (t)= \omega_0+\omega_1 t$ \cite{Mukhopadhyay2007}. 
In the present work, let us suppose the structure of $\omega$ 
as follows 
\begin{eqnarray}
\omega(t)=\omega_0+\frac{\omega_1\tau}{t^2},\label{omega}
\end{eqnarray}
where $\omega_0$ and $\omega_1$ are two constants and $\tau$ is a
parameter having dimension of $t^2$. Here $\tau$ apart from being
a non-negative mathematical parameter, rather bears some deeper physical
significance. It has been shown that $\tau$ represents the time-scale of 
evaporation of Bose-Einstein condensates which include a time-dependent
$\Lambda$ \cite{Dymnikova1998,Dymnikova2000,Dymnikova2001}. In this context 
it is interesting to note that Mukhopadhyay et al. \cite{Mukhopadhyay2010} 
consider $\tau$ in their equation of state in the form $\tau = t/(1+\omega)^{1/n}$. 
Thus, they argue that for the present dust-filled Universe ($\omega=0$),
$\tau$ is equal to $t$ whereas for vacuum fluid $\tau$ becomes
meaningless. However, in the present case the expression for $\omega$ 
immediately suggests that for very small $t$, i.e. in the early Universe, 
the contributions from the second term, $\omega_1\tau/t^2$, in the 
right hand side of equation (\ref{omega}) were significant. But for large $t$, 
i.e. for late Universe, this term tends to zero and consequently $\omega(t)$
converges to a nearly constant value $\omega_0$. This may be one of the
reasons that at present we can not distinguish between a constant 
and a variable equation of state as mentioned in the Introduction.

By the use of the equation (\ref{omega}) in equation (\ref{H2}) and solving the
resulting equation we get our solution set as
\begin{eqnarray}
a(t) = \left[(1+\omega_0)t^2-\omega_1\tau\right
]^{\frac{1}{3(1-\alpha)(1+\omega_0)}},\label{sol1}
\end{eqnarray}
\begin{eqnarray}
H(t) = \frac{2t}{3(1-\alpha)[(1+\omega_0)t^2-\omega_1\tau]},\label{sol2}
\end{eqnarray}
\begin{eqnarray}
\rho(t) = \frac{t^2}{6\pi G
(1-\alpha)[(1+\omega_0)t^2-\omega_1\tau]^2},\label{sol3}
\end{eqnarray}
\begin{eqnarray}
\Lambda(t) = \frac{4\alpha
t^2}{3(1-\alpha)[(1+\omega_0)t^2-\omega_1\tau]^2}.\label{sol4}
\end{eqnarray}

Now, since the Universe is at present old enough, one can easily neglect the terms involving
powers of $1/t^2$ greater than one. Thus the equations (\ref{sol2})-(\ref{sol4}) transform to
\begin{eqnarray}
H(t) =\frac{2}{3(1-\alpha)(1+\omega_0)t}\left[1+\frac{\omega_1\tau}{(1+\omega_0)t^2}\right],\label{sol2a}
\end{eqnarray}
\begin{eqnarray}
\rho(t) = \frac{1}{6\pi G
(1-\alpha)(1+\omega_0)^2t^2}\left[1+\frac{2\omega_1\tau}{(1+\omega_0)t^2}\right],\label{sol3a}
\end{eqnarray}
\begin{eqnarray}
\Lambda(t) = \frac{4\alpha}{3(1-\alpha)^2(1+\omega_0)^2
t^2}\left[1+\frac{2\omega_1\tau}{(1+\omega_0)t^2}\right].\label{sol4a}
\end{eqnarray}

From equation (\ref{sol2a}) it is clear that for large $t$, the term
$\omega_1\tau/(1+\omega_0)t^2$ is negligibly small and hence we
get the relationship $H\propto t^{-1}$. Similarly from equations
(\ref{sol3a}) and (\ref{sol4a}) neglecting the terms involving $t^{-4}$ one can
arrive at the relations $\rho\propto t^{-2}$ and $\Lambda \propto
t^{-2}$. Thus for large $t$ we can recover the relationships
$H\propto t^{-1}$, $\rho\propto t^{-2}$ and $\Lambda \propto
t^{-2}$ obtained by Ray et al. \cite{Ray2007b} for constant
$\omega$. This proves the generality of the present work so far as
the time variations of $H$, $\rho$ and $\Lambda$ are concerned.

\section{Physical Features of the Parameters}

Using equations (\ref{sol2a})-(\ref{sol4a}) we can obtain the expressions for the
matter-energy density $\Omega_m$ and dark-energy density
$\Omega_{\Lambda}$ as
\begin{eqnarray}
\Omega_m \equiv \frac{8\pi G \rho}{3H^2} = (1-\alpha)
\left[1-\frac{4\omega_1^2\tau^2}{(1+\omega_0)^2t^4}\right],\label{matt}
\end{eqnarray}
\begin{eqnarray}
\Omega_{\Lambda} \equiv \frac{\Lambda}{3H^2} =
\alpha\left[1-\frac{4\omega_1^2\tau^2}{(1+\omega_0)^2t^4}\right].\label{dark}
\end{eqnarray}

Adding equations (\ref{matt}) and (\ref{dark}) we can obtain
\begin{eqnarray}
\Omega = \Omega_m + \Omega_{\Lambda} =
1-\frac{4\omega_1^2\tau^2}{(1+\omega_0)^2t^4}.\label{total}
\end{eqnarray}

\begin{figure}
\begin{center}
\includegraphics[angle=0,width=9cm]{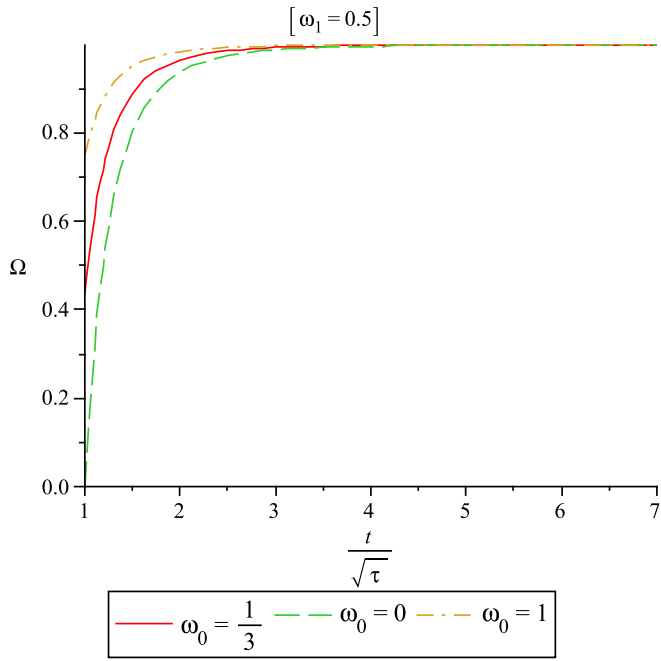}
\caption{Variation of density parameter with cosmic time for the specified values of $\omega_0$ when
$\omega_1 = 0.5$ (Equation 18).}
 \label{fig1}
\end{center}
\end{figure}

From equations (\ref{matt}) and (\ref{dark}) we find that for large $t$, $\Omega_m
\simeq (1-\alpha)$ and $\Omega_{\Lambda}\simeq \alpha$. For the
same $\Lambda$ model with constant $\omega$, Ray et al.
\cite{Ray2007b} obtained $\Omega_m =(1-\alpha)$ and
$\Omega_{\Lambda}= \alpha$. So, in this case also, we find that
the present work is a generalization of the work of Ray et al.
\cite{Ray2007b}. Further from equation (\ref{total}) we get 
$\Omega \simeq 1$ when $t$ is very large. This result is
also compatible with the observational results and the work of Ray
et al. \cite{Ray2007b}. Moreover, since the second term in the
right hand side of equation (\ref{total}) is clearly positive, 
$\Omega < 1$. But, the sum of all types of energy densities
being equal to $1$, the term $4\omega_1^2\tau^2/(1+\omega_0)^2t^4$
in equation (\ref{total}) can be interpreted as the sum of energy densities
due to radiation and curvature, or in absence of any curvature,
the energy density coming from radiation only. It has already been
mentioned that for large $t$, the term
$4\omega_1^2\tau^2/(1+\omega_0)^2t^4$ is very small. This result
is compatible with the idea that at present both radiation-energy
density and energy density due to curvature are negligibly small
and almost entire energy density of the Universe comes from
matter- and dark-energy densities. In this context we would like 
to mention the work of Mukhopadhyay et al. \cite{Mukhopadhyay2009} 
where, for the $\Lambda \propto H^2$ model with constant $\omega$,
it has been shown that the sum of matter- and dark energy-densities 
is equal to $1$ for both early and present Universe. But this result 
$\Omega_m + \Omega_{\Lambda} = 1$ is valid for the late Universe also only when 
the Universe is composed of two fluids having barotropic indices $\omega_a$ and
$\omega_b$. In the present work, however, without resorting to any
special assumption, it has been possible to show that with the
passage of time the density parameter $\Omega~(=\Omega_m + \Omega_{\Lambda})$ 
smoothly moves from values less than $1$ to the currently accepted value $1$. 
This aspect is clear from the plot for $\Omega$ vs $t$ in the Fig. 1, where 
variation of density parameter with cosmic time for the specified values of $\omega_0$ 
when $\omega_1 = 0.5$ has been shown. 

Let us now calculate the deceleration parameter to see it's physical status. 
For this, by using equation (\ref{sol2}), we get the expression for the deceleration
parameter $q$ as
\begin{eqnarray}
q =\frac{3(1-\alpha)}{2}\left[(1+\omega_0)+\frac{\omega_1\tau}{t^2}\right]-1.\label{q1}
\end{eqnarray}

From equation (\ref{q1}) we find that for small $t$, the term
$\omega_1\tau/t^2$ had a significant contribution and consequently
the Universe was in a decelerating phase. But with the passage of
time the term $\omega_1\tau/t^2$ became negligibly small and hence
the Universe switched over to the present accelerating epoch. This
clearly shows the signature flipping of $q$ which is essential for
the $\Lambda$-CDM cosmology. Therefore, for large $t$ one can neglect the second
term inside the bracket in equation (\ref{matt}), when the equation (\ref{q1}) 
reduces to to following form as
\begin{eqnarray}
q =\frac{3\Omega_m}{2}\left[(1+\omega_0)+\frac{\omega_1\tau}{t^2}\right]-1.
\end{eqnarray}

\begin{figure}
\begin{center}
\includegraphics[angle=0,width=9cm]{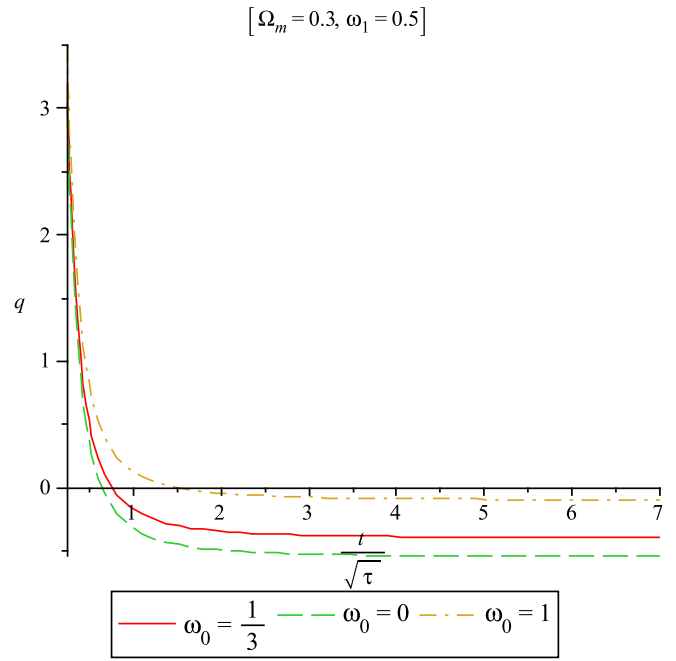}
\caption{Variation of deceleration parameter with cosmic time for the specified values of $\omega_0$ when
$\omega_m=0.3$ and $\omega_1 = 0.5$ (Equation 20).}
 \label{fig2}
\end{center}
\end{figure}

Now, assuming the value $\Omega_m=0.30$, which is within the
range $\Omega_m=0.330 \pm 0.035$
\cite{Vishwakarma2002,Turner2002,Rebolo2003,Alcaniz2004} and
$\omega_0=0$, the present value of the $q$ comes out as
$q_0=-0.55$. This value of $q_0$ nicely fits with the presently
accepted value of that parameter as $q_0= -0.50 \pm 0.05$ for 
an accelerating Universe \cite{Tripp1997,Sahni1999} (Fig. 2). This value of
$q_0$ was earlier obtained by Ray et al \cite{Ray2007b} for
$\omega=0$. Thus through the present work, it has been possible to
recover the value of $q_0$ as obtained by Ray et al. \cite{Ray2007b}.

\begin{figure}
\begin{center}
\includegraphics[angle=0,width=9cm]{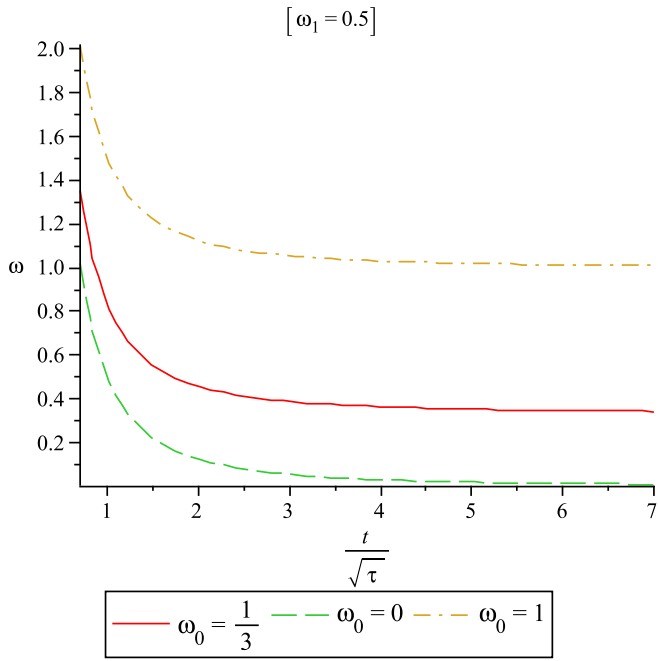}
\caption{Variation of equation of state parameter with cosmic time for the specified values of $\omega_0$ when
$\omega_1 = 0.5$ (Equation 8).}
 \label{fig3}
\end{center}
\end{figure}

In the same spirit we would like to look at the equation of state 
expressed in the equation (9) and plot the variation of equation of 
state parameter with cosmic time. The evolution of $\omega$ are shown 
here for radiation, dust, and stiff fluid phase with solid, dashed and chain-dashed 
lines respectively (Fig. 3). For all these, $\omega_1$ is taken to be $0.5$. 
However, it is observed that the nature of plots related to $\omega$ 
does not depend on the values of $\omega_1$.

\section{Discussion}

In the present investigation some interesting physics are explored 
by assuming a particular time-dependent form of the equation of state parameter
$\omega$ to track down the time evolution of the Universe. It is shown that 
at present both radiation-energy density and energy density due to curvature 
are negligibly small such that almost the entire energy density of the Universe 
comes from matter- and dark-energy densities. With the passage of time 
the cosmic density parameter $\Omega$ smoothly moves from values less than $1$ 
to the currently accepted value $1$ as shown in the Fig. 1. 

It has also been possible to recover the results of 
Ray et al. \cite{Ray2007b} so far as the expressions of $H$, $\rho$,
$\Lambda$, $\Omega_m$, $\Omega_{\Lambda}$ and $q$ are concerned.
In particular, the expression for $q$ shows the required signature
change during cosmic evolution proving that the present
acceleration of the Universe is a recent phenomena (Fig. 2). The calculated
value of $q$ is compatible with the modern accepted value of this parameter.
The present investigation has also been interesting to show a glimpse of
the small amount of radiation-energy density left over from the
early Universe in the form of CMBR and tiny fraction of energy
density, if it exists at all at present, due to curvature of
space. This justifies the comment made by Mukhopadhyay et al.
\cite{Mukhopadhyay2009} that we are still not in a position to
rule out $k=\pm 1$ cosmologies. 

 It is to note that the present work has also been 
successful in justifying the idea of Ray et al \cite{Ray2007b} that 
time-dependent equation of state parameter is essential for studying 
the complete time-evolution of the cosmos (Fig. 3). In connection to the work of 
Mukhopadhyay and Ray \cite{Mukhopadhyay2005} it is to be mentioned here that for quintessence, 
vacuum fluid and phantom energy, the rate of acceleration should be higher, e.g. 
for $\omega=-0.5$, $-1.0$ and $-2.0$ the resultant deceleration parameter 
values are respectively, $q=-0.775$, $-1.0$ and $-1.45$. On the other
hand, for stiff fluid ($\omega=1.0$), $q=-0.1$. This suggest that more smaller
the value of $\omega$ the higher is the rate of acceleration resulting the 
so-called Big Rip \cite{Caldwell2003} or Partial Rip \cite{Stefancic2004} scenario
due to divergence of scale factor. One can also note the work of 
Kuhlen et al. \cite{Kuhlen2005} where they consider a range of parameter space: 
$\omega=-0.5$, $-0.75$, $-1.0$, $-1.25$ and $-1.5$ for their simulations 
to reproduce the time-evolution of the Universe. However, in the present 
investigation by considering an equation of state in the form 
$\omega(t)=\omega_0+\frac{\omega_1\tau}{t^2}$ we are able to get 
the total range of time evolution of the Universe. Here, as can be seen 
from the Fig. 3 that starting from different values for radiation, dust and stiff fluid 
all the situations ultimately approach to the same physical feature.

Finally, we would like to mention that in the present model $\rho$ is positive 
and hence the Weak Energy Condition (WEC) as well as the Null Energy Condition (NEC) are obeyed here.

\section*{Acknowledgments} The authors (FR \& SR) would like to
express their gratitude to the authority of IUCAA, Pune, India for
providing them Visiting Associateship under which a part of
this work was carried out.


\begin{thebibliography}{0}


\bibitem{Riess et al.1998} A. G. Riess et al., Astron. J. {\bf 116}, 1009 (1998)

\bibitem{Perlmutter et al.1998} S. Perlmutter et al., Nat. {\bf 391}, 51 (1998)

\bibitem{Barrow1999} J. D. Barrow and J. Maguijo, astro-ph/9907354 (1999)

\bibitem{Garcia-Berrow1999} E. Garcia-Berrow et al., astro-ph/9907440 (1999).

\bibitem{Celerier2000} M. N. C{'e}l{'e}rier, Astron. Astrophys. {\bf 353}, 63 (2000)

\bibitem{Overduin1998} J. M. Overduin  and F. I. Cooperstock, Phys. Rev. D {\bf
58}, 043506 (1998)

\bibitem{Sahni2000} V. Sahni and A. Starobinsky, Int. J. Mod. Phys. D {\bf 9}, 373 (2000) 

\bibitem{Ray2007a} S. Ray, U. Mukhopadhyay and S. B. Dutta Choudhury, Int. J. Mod. Phys.D {\bf 16}, 1791 (2007)

\bibitem{Knop2003} R. A. Knop et al., Astrophys. J. {\bf 598}, 102 (2003)

\bibitem{Tegmark2004} M. Tegmark et al., Astrophys. J. {\bf 606}, 702 (2004)

\bibitem{Chevron2000} S. V. Chevron and V. M. Zhuravlev, Zh. Eksp. Teor. Fiz. {\bf 118}, 259 (2000)

\bibitem{Zhuravlev2001} V. M. Zhuravlev, Zh. Eksp. Teor. Fiz. {\bf 120}, 1042 (2001)

\bibitem{Peebles2003} P. J. E. Peebles and B. Ratra, Rev. Mod. Phys. {\bf 75}, 559 (2003)

\bibitem{Kujat2002} J. Kujat et al., Astrophys. J. {\bf 572}, 1 (2002)

\bibitem{Bartelmann2005} M. Bartelmann et al., New Asrton. Rev. {\bf 49}, 199 (2005)

\bibitem{Huterer2001} D. Huterer and M. S. Turner,  Phys. Rev. D {\bf 64}, 123527 (2001)

\bibitem{Weller2002} J. Weller and A. Albrecht, Phys. Rev. D {\bf 65}, 103512 (2002)

\bibitem{Polarski2001} D. Polarski and M. Chavellier, Int. J. Mod. Phys. {\bf 10}, 213 (2001)

\bibitem{Linder2003a} E. V. Linder, Phys. Rev. Lett. {\bf 90}, 091301 (2003)

\bibitem{Linder2008} E. V. Linder, Gen. Rel. Gravit. {\bf 40}, 329 (2008)

\bibitem{Rapatti2005} D. Rapatti, S. W. Allen and J. Weller, Mon. Not. R. Astron. Soc. {\bf 360}, 555 (2005)

\bibitem{Corasaniti2003} P.-S. Corasaniti and E. J. Copeland, Phys. Rev. D {\bf 67}, 063521 (2003)

\bibitem{Linder2005} E. V. Linder and D. Huterer, Phys. Rev. D {\bf 72}, 043509 (2005)

\bibitem{Linder2003b} E. V. Linder in {\it Identification of dark matter (IDM 2002)} {\bf p. 52}, astro-ph/0210217 (2002)

\bibitem{Ratra1988} B. Ratra and P. J. E. Peebles, Phys. Rev. D {\bf 37}, 3406 (1988)

\bibitem{Turner1997} M. S. Turner and M. White, Phys. Rev. D {\bf 56}, R4439 (1997)

\bibitem{Caldwell1998} Caldwell et al.,  Phys. Rev. Lett. {\bf 80}, 1582 (1998)

\bibitem{Liddle1999} A. R. Liddle and R. J. Scherrer,  Phys. Rev. D {\bf 59}, 023509 (1999)

\bibitem{Steinhardt1999} P. J. Steinhardt et al., Phys. Rev. D {\bf 59}, 123504 (1999)

\bibitem{Bhui2005} B. Bhui, B. C. Bhui and F. Rahaman, Astrophys. Space Sc. {\bf 299}, 61 (2005)

\bibitem{Rahaman2006} F. Rahaman, B. Bhui and B. C. Bhui, Astrophys. Space. Sc. {\bf 301}, 47 (2006)

\bibitem{Rahaman2009} F. Rahaman, M. Kalam and S. Chakraborty, Acta Phys. Polon. B {\bf 40}, 25 (2009)

\bibitem{Amendola2003} L. Amendola, Mon. Not. R. Astron. Soc. {\bf 342}, 221 (2003)

\bibitem{Padmanabhan2003} T. Padmanabhan and T. Roychowdhury, Mon. Not. R. Astron. Soc. {\bf 344}, 823 (2003)

\bibitem{Riess2001} A. G. Riess, Astrophys. J. {\bf 560}, 49 (2001)

\bibitem{Mukhopadhyay2007} U. Mukhopadhyay, P. P. Ghosh, M. Khlopov and S. Ray, astro-ph/0711.0686 (2007)

\bibitem{Mukhopadhyay2008a} U. Mukhopadhyay, S. Ray and S. B. Dutta Choudhury, Int. J. Mod. Phys. D {\bf 17}, 301 (2008)

\bibitem{Usmani2008} A. A. Usmani, P. P. Ghosh, U. Mukhopadhyay, P. C. Ray and S. Ray, Mon. Not. R. Astron. Soc. {\bf 386}, L
92 (2008)

\bibitem{Mukhopadhyay2005} U. Mukhopadhyay and S. Ray, N. B. Math. {\bf II}, 51 (2009)

\bibitem{Mukhopadhyay2008b} U. Mukhopadhyay, S. Ray and A. A. Usmani, gr-qc/0811.0782 (2008)

\bibitem{Ray2007b} S. Ray, U. Mukhopadhyay and Xin He-Meng, Gravit. Cosmol. {\bf 13}, 142 (2007)

\bibitem{Gliner1965} F. Gliner, ZETF {\bf 49}, 542 (1965)

\bibitem{Majernik2001} V. Majernik, Phys. Lett. A {\bf 282}, 362 (2001)

\bibitem{Vishwakarma2001} R. G. Vishwakarma, Gen. Relativ. Gravit. {\bf 33}, 1973  (2001)

\bibitem{Shapiro2005} I. L. Shapiro, J. Sol{\`a} and H.\u{S}tefan\u{c}i{\'c}, J. Cosmol. AstroParticle. {\bf 01}, 012 (2005)

\bibitem{Vereschagin2006} G. V. Vereschagin and G. Yegorian, Class. Quantum Gravit. {\bf 23}, 5049 (2006)

\bibitem{Gurzadyan2003} V. G. Gurzadyan and S. -S. Xue, Mod. Phys.Lett. A {\bf 18}, 561 (2003)

\bibitem{Beesham1993} A. K. Beesham, Phys. Rev. D {\bf 48}, 8 (1993)

\bibitem{Carvalho1992} J. C. Carvalho et al., Phys. Rev. D {\bf 46}, 2404 (1992)

\bibitem{Waga1993} I. Waga, Astrophys. J. {\bf 414}, 436 (1993)

\bibitem{Lima1994} J. A. S. Lima and J. C. Carvalho, Gen. Relativ. Gravit. {\bf 26}, 909 (1994)
.
\bibitem{Vishwakarma2002} R. G. Vishwakarma, Class. Quantum Gravit. {\bf 19}, 4747 (2002)

\bibitem{Cohen1977} A. G. Cohen, D. B. Kaplan and A. E. Nelson, Phys. Rev. Lett. {\bf 82}, 4971 (1999)

\bibitem{Hsu2004} S. D. H. Hsu, Phys. Lett. B {\bf 594}, 13 (2004)

\bibitem{Mukhopadhyay2010} U. Mukhopadhyay, S. Ray and F. Rahaman, Int. J. Mod. Phys. D {\bf in press}, (2010)

\bibitem{Dymnikova1998} I. Dymnikova and M. Khlopov, Grav. Cosmol. Suppl. {\bf 4}, 50 (1998)

\bibitem{Dymnikova2000} I. Dymnikova and M. Khlopov, Mod. Phys. Lett. A {\bf 15}, 2305 (2000)

\bibitem{Dymnikova2001} I. Dymnikova and M. Khlopov, Eur. Phys. J. C {\bf 20}, 139 (2001)

\bibitem{Mukhopadhyay2009} U. Mukhopadhyay, P. C. Ray, S. Ray and S. B. Dutta Choudhury, Int. J. Mod. Phys. D {\bf 18}, 389 (2009)

\bibitem{Turner2002} M. S. Turner, Astrophys. J. {\bf 576}, L101 (2002)

\bibitem{Rebolo2003} R. Rebolo, Nucl. Phys. B (Proc. Suppl.) {\bf 114}, 3 (2003)

\bibitem{Alcaniz2004} J. S. Alcaniz, Phys. Rev. D {\bf 69}, 083521 (2004)

\bibitem{Tripp1997} R. Tripp, Astron. Astrophys. {\bf 325}, 871 (1997)

\bibitem{Sahni1999} V. Sahni, Pramana {\bf 53}, 937 (1999)

\bibitem{Mukhopadhyay2005} U. Mukhopadhyay and S. Ray, astro-ph/0510557 (2005)

\bibitem{Caldwell2003} R. R. Caldwell, M. Kamionkowski and N. N. Weinberg, Phys. Rev. Lett. {\bf 91}, 071301 (2003)

\bibitem{Stefancic2004} H. \u{S}tefan\u{c}i{\'c}, Phys. Lett. B {\bf 595}, 9 (2004)

\bibitem{Kuhlen2005} M. Kuhlen et al., Mon. Not. R. Astron. Soc. {\bf 357}, 387 (2005)



\end{thebibliography}
\end{document}